# MRI-based Surgical Planning for Lumbar Spinal Stenosis

**Gabriele Abbati**      GABB@ROBOTS.OX.AC.UK
*Department of Engineering Science*
*University of Oxford*
*Eagle House, Walton Well Road, Oxford OX2 6ED, United Kingdom*

**Peter J. Schüffler**      SCHUEFFP@MSKCC.ORG
*Thomas Fuchs Lab, Medical Machine Learning & Computational Pathology*
*Memorial Sloan Kettering Cancer Center*
*417 E 68th St., New York, NY 10065 (USA)*

**Sebastian Winklhofer**      SEBASTIAN.WINKLHOFER@USZ.CH
*Klinik für Neuroradiologie*
*University Hospital Zürich*
*Rämistrasse 100, 8091 Zürich, Switzerland*

**Jakob M. Burgstaller**      JAKOB.BURGSTALLER@USZ.CH
**Ulrike Held**      ULRIKE.HELD@USZ.CH
**Johann Steurer**      JOHANN.STEURER@USZ.CH
*Horten Centre for Patient Oriented Research and Knowledge Transfer*
*University Hospital Zürich*
*Pestalozzistrasse 24, 8032 Zürich, Switzerland*

**Stefan Bauer**      STEFAN.BAUER@INF.ETHZ.CH
**Joachim M. Buhmann**      JBUHMANN@INF.ETHZ.CH
*Department of Computer Science*
*ETH Zürich*
*Universittstrasse 6, 8006 Zürich, Switzerland*

## Abstract

The most common reason for spinal surgery in elderly patients is lumbar spinal stenosis (LSS). For LSS, treatment decisions based on clinical and radiological information as well as personal experience of the surgeon shows large variance. Thus a standardized support system is of high value for a more objective and reproducible decision. In this work, we develop an automated algorithm to localize the stenosis causing the symptoms of the patient in magnetic resonance imaging (MRI). With 22 MRI features of each of five spinal levels of 321 patients, we show it is possible to predict the location of lesion triggering the symptoms. To support this hypothesis, we conduct an automated analysis of labeled and unlabeled MRI scans extracted from 788 patients. We confirm quantitatively the importance of radiological information and provide an algorithmic pipeline for working with raw MRI scans.

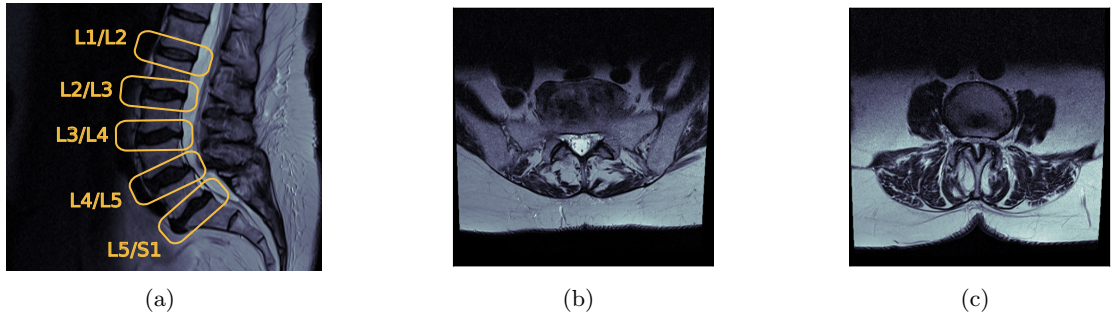

Figure 1: Examples of T2-weighted MRI. **1(a)** The five segments are highlighted yellow in a sagittal scan. **1(b)** Axial scan of a patient without symptoms and without narrowing of the spinal channel (white spot in the center). **1(c)** Example with extreme narrowing.

## 1. Introduction

The lumbar spine consists of the five vertebrae (*levels* or *segments*) L1–L5. The vertebral discs connect adjacent levels and are denoted as L1/L2, L2/L3, L3/L4, L4/L5, L5/S1, where S1 is the first vertebra of the underlying *sacral region* (see figure 1(a)). Lumbar Spinal Stenosis (LSS) is the most common indicator for spine surgery in patients older than 65 years (Deyo (2010)). The North American Spine Society defines LSS as "[...] diminished space available for the neural and vascular elements in the lumbar spine secondary to degenerative changes in the spinal canal [...]" (Kreiner et al. (2014)). Symptoms such as gluteal and/or lower extremity pain and/or fatigue might occur, possibly associated with back pain. Magnetic resonance imaging (MRI, illustrated in figs. 1(b) and 1(c)) and the patient's clinical course contribute to diagnosis and treatment formulation. When conservative treatments such as physiotherapy or steroid injections fail, decompression surgery is frequently indicated (Deyo (2010)). Depending on the clinical presentation of the patient and corresponding imaging findings, surgeons decide which segments to operate. This decision process exhibits wide variability (Weinstein et al. (2006); Irwin et al. (2005)), while associations between imaging and symptoms are still not entirely clear (Jensen et al. (1994); Burgstaller et al. (2016)). These issues motivate the search for objective methods to help in surgery planning. Since the definition of LSS implies anatomic abnormalities, MRI plays a fundamental role in diagnosis (Steurer et al. (2011)). Andreisek et al. (2014) identified 27 radiological criteria and parameters for LSS. However, correlations between imaging procedures, clinical findings and symptoms is still unclear, and research efforts show contradictory results (Haig et al. (2006); Ishimoto et al. (2013)).

This paper comprehensively determines the important role of radiological parameters in LSS surgery planning, in particular by modeling surgical decision-making: to the best of our knowledge, no machine learning approach has been applied in this direction before. In section 2, we automatically predict surgery locations with 22 manual radiological features comparing five different classifiers. We obtain accuracies of 85.4% using random forests and show features associated with stenosis are commonly chosen by all classifiers. In section 3, the highly heterogeneous MRI dataset is preprocessed and a convolutional neural network and convolutional autoencoder are trained to accomplish the same task as before, without any knowledge of the underlying structure of LSS. The automatic preprocessing of raw MRI scans is a key contribution of this work and code with examples will be released in the final version. Both algorithms achieve accuracies of 69.8% and 70.6%, respectively, in mimicking surgeons' decisions, showing the high relevance of radiological features in LSS treatment. Finally, we conclude with a discussion in section 4.

## 2. Surgical Prediction from Numerical Dataset

### 2.1 The Numerical Dataset

Radiological T1-weighted and T2-weighted scans from 788 LSS patients have been collected in a multi-center study by Horten Zentrum (Zürich, CH). For every segment and patient, radiologists manually scored 6 quantitative features (e.g. area of spinal canal in $mm^2$) and 16 qualitative features (e.g. severity grade of compromise of a given vertebral region) known to be most relevant for assessing stenosis (Andreisek et al. (2014)), forming the "numerical" dataset. A description of the features can be found in the Supplement (A.1).
431 of 788 patients underwent surgery. The Numeric Rating Scale (NRS, Downie et al. (1978)) for pain assessment was employed to understand whether the intervention improved a certain patient's condition or not. NRS differences larger than 2 points before and six months after surgery were considered as improvement, as failure otherwise. In total, 321 of 431 patients exhibited improvement of NRS after surgery. As there is no information gain from unsuccessful operations, the following analysis addresses the subset of the 321 improved patients, yielding a total of 1385 segments as data points.

### 2.2 Methods

We consider every segment independently as a data vector **x** consisting of its 22 feature values. The target is represented by a binary variable $y$ (to operate / not to operate). This binary classification framework is tackled with the following algorithms: $K$-nearest neighbors (KNN), linear discriminant analysis (LDA), quadratic discriminant analysis (QDA), support vector machine (SVM), and random forest (RF). Implementations from the *scikit-learn* (Pedregosa et al. (2011)) library are employed. The area under the receiver operating characteristic (ROC) curve is a natural choice for evaluating binary classifiers' performances, and it is combined with 20-fold cross validation.
To evaluate the influence of individual features, forward selection and backward selection are employed to choose the best 3, 5, 8, 10, 12, 15 and 18 features: with 5 different classifiers, a single feature can be chosen for a total of 70 times (7 sets × 5 classifiers × 2 algorithms). Thus we can evaluate how often a feature is considered to be among the most relevant ones for surgery prediction. This procedure is again validated through 20-fold cross validation.

### 2.3 Results

For parameter-optimized binary classifiers, box plots describing the area under the ROC curve (AUC) obtained with 20-fold cross validation are shown in figure 2(a). The best results are achieved with an optimized random forest classifier: the mean over the AUC returned by the cross validation is 85.4%, with a standard deviation of 3.26%. The precision obtained here is particularly significant if we consider the relatively low agreement rates between doctors in determining treatments for LSS (Lurie et al. (2008); Fu et al. (2014)). Feature selection indicates that `SegCentralZone` (assesses the compromise of the central zone of the vertebra), `SegCSarea` (area of the section of the spinal cord in mm$^2$) and `SegFluidSign` (relation from fluid to cauda equina) as the most important features for assessing stenosis: these are chosen in 88.57%, 87.14% and 70.00%, respectively, of the total trials with feature selection algorithms (total ranking in figure 2(b)). All three features are known to be strongly related to spinal stenosis (Andreisek et al. (2014)). The results show that radiological data actually helps in assessing LSS and planning surgical treatments.

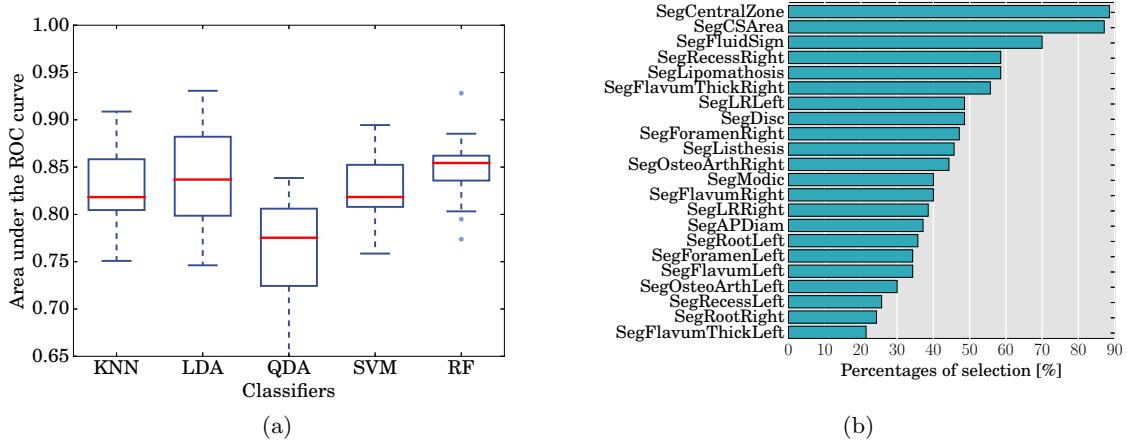

Figure 2: Summary of the classifiers for segmental surgery prediction. **2(a)** The box plots of the 20-fold cross validation. All classifiers show a strong signal between radiological data and surgical treatments. **2(b)** Feature ranking as described in the text. The three most important features are `SegCentralZone`, `SegCSarea` and `SegFluidSign`.

## 3. Surgical Prediction from Radiological Images

Fully automated MRI-based surgery planning would be a helpful tool, as it can substantially speed up the process by skipping manual scoring while reducing the variability of human assessment. Therefore, we aim to directly learn features from raw MRI scans.

### 3.1 The Image Dataset

The above described dataset of 788 LSS patients contains a great variety of T1-weighted and T2-weighted sagittal, coronal and axial series scans (see figure 3 for four typical examples). Since the images come from seven different institutions, the dataset is heterogeneous: not all types of MRI scans listed above are always available, and often only a small subset of the segments is accessible. Further, different machines vary in resolution (from 320x320 to 1024x1024 pixels) and scanning frequency (0.2 to 1 scan/mm).

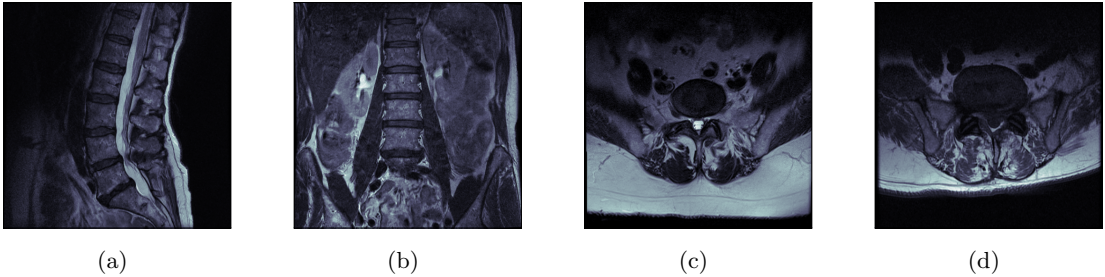

(a) (b) (c) (d)

Figure 3: Typical examples of the different MRI scans: figs. 3(a) to 3(c) T2-weighted (3(a), sagittal; 3(b), coronal; 3(c), axial); 3(d) T1-weighted axial.

To keep the same segment-wise approach as before, we decide to employ only the T2-weighted axial scans (e.g. figure 3(c)), as they picture the whole lumbar spine and can be easily chopped into single segment sub-series. T2-weighted imaging pictures the spinal canal white in contrast to T1-weighted images, in which the canal is dark and hardly visible (3(d)). Further, T2-weighted axial scans are the most common series in the dataset. The image dataset includes the same 321 operated patients with improved NRS.

### 3.2 Image Preprocessing & Data Augmentation

All images are cropped and resized to 128x128 pixels, in order to keep the central section. Because of the various scanning frequencies, we then linearly interpolate to a desired number of equally spaced slices: to sufficiently describe the vertebral disc, yet keep the data structure simple, we use four subimages for each segment. **We employed following data augmentation**: rotation by a random angle $\alpha \in [-10°; 10°]$; sagittal mirroring; inversion of the order of the slides (since the MRI machine can scan upwards or downwards); application of random Gaussian noise (zero-mean and 5% standard deviation); random brightness alteration (maximum alteration at 5%). Each image is augmented 20 times by this pipeline, each time every augmentation technique is randomly applied or not applied.

### 3.3 Methods

Deep learning algorithms have already shown great success in a variety of image recognition problems (Krizhevsky et al. (2012); Farabet et al. (2013)). Convolutional Neural Networks (CNN, implementation details can be found in LeCun et al. (1998)) are image processing algorithms that are able to extract image features regardless of their position, which is especially useful in our case since scans are not always optimally centered on the spine. Due to the small sample size, a simple architecture is needed to prevent overfitting. **Our CNN has the following structure:** first convolutional layer (filters size 5x5, 128 masks), followed by a max-pooling layer; second convolutional layer (filters size 5x5, 64 masks), followed by a max-pooling layer; a fully connected layer, 2048 nodes; a further fully connected layer, 1024 nodes. Rectifier Linear Units (ReLU) are a common choice for this kind of network. The network structure is illustrated in fig 4, step 3. The cost function minimized during training is the mean of the softmax cross-entropy function between the output $\mathbf{x}$ and the actual label vector $\mathbf{z}$, $\mathcal{L} = -\mathbf{z} \log \sigma(\mathbf{x}) - (1-\mathbf{z}) \log [1 - \sigma(\mathbf{x})]$, where $\sigma(\mathbf{x})$ is the softmax function. The optimizer used for the minimization is AdaGrad (Duchi et al. (2011)). Implementation is done in Python using TensorFlow (Abadi et al. (2016)).

The major inherent vice in this approach is the need of labeled examples. We learn from 1576 labeled scanned segments from 321 successfully operated patients. On the other hand, if we were able to include unlabeled segments in the analysis, we could take advantage of all 4031 segments from the 788 patients.

Unsupervised learning methods do not need labeled examples. The autoencoder algorithm (Zemel (1994)) is used to reduce the dimensionality of the problem: it consists of an encoder function $\mathbf{h} = f(\mathbf{x})$ and a decoder function $\mathbf{r} = g(\mathbf{h})$. The autoencoder is trained to copy the input to the output, but it is not given the resources to do so exactly (undercompleteness property). In this way an approximation of the input is returned and the model is forced to prioritize the most relevant aspects of the input. As the autoencoder does not need labels for the surgery, all 4031 segments can be used. **An autoencoder sufficient for our needs** can be built by mirroring the CNN and learning how to "invert" the convolutional and the max pooling layers (Zeiler et al. (2010)) into *deconvolutional layers*: first convolutional layer (filters size 5x5, 128 masks), followed by a max-pooling layer; second convolutional layer (filters size 5x5, 64 masks), followed by a max-pooling layer; a fully connected layer, 1024 nodes; a fully connected layer, 128 nodes (*bottleneck*); a fully connected layer, 1024 nodes; first unpooling and deconvolutional layer (filters size 5x5, 64 masks); second deconvolutional layer (filters size 5x5, 128 masks). This autoencoder reconstructs the original 3D image, and in the middle layer (the bottleneck), we find a 128-number code that identifies each image sufficiently for its reconstruction. We train the autoencoder on all unlabeled images to minimize the difference tensor $\mathbf{J} = (\mathbf{X}_{\text{orig}} - \mathbf{X}_{\text{reconstr}})^2$, where $\mathbf{X}_{\text{orig}}$ is the original image and $\mathbf{X}_{\text{reconstr}}$ is its reconstruction. After training, the autoencoder is used to encode all labeled images and their 128-number codes are used as features in the same classification experiments as in section 2.

### 3.4 Results

The complete pipeline from the MRI preprocessing to the surgery classification is depicted in fig. 4. For both CNN and autoencoder, the available image datasets are split into training and test set with a 80/20 ratio. The training sets are augmented as previously described and the networks are trained for 100 epochs. Learning curves are available in the Supplement (A.2). On the test set, the CNN reaches an AUC of 69.8%. This is significantly lower than the AUC obtained with the numerical dataset, but it is still confirming the existence of a signal in the MRI images, and enforces the idea that radiological data are linked to stenosis diagnosis and treatment. Considering the small size of the training data, we are confident that higher precisions can be obtained if the present dataset is improved and expanded. The autoencoder learns successfully to reconstruct the images (fig. 5). While some details are missed, it is noticeable that the dimension of a picture is now extremely reduced from $128 \times 128 \times 4 = 65536$ numbers to 128. When training and testing the binary classifiers from section 3 with the codes from the labeled segments, the highest mean AUC for a 20-fold cross validation test is given by a optimized LDA classifier, at 70.6%, with a corresponding standard deviation of 6.69%. The mild improvement can be explained by the extension of the dataset to the non-labeled segments.

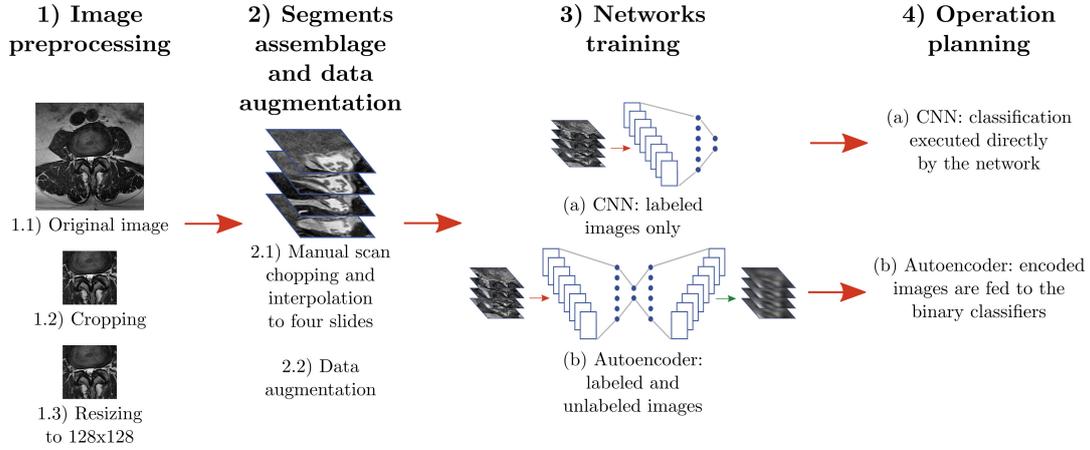

Figure 4: Proposed computing pipeline from preprocessing of raw MRI pictures to learning of surgical planning

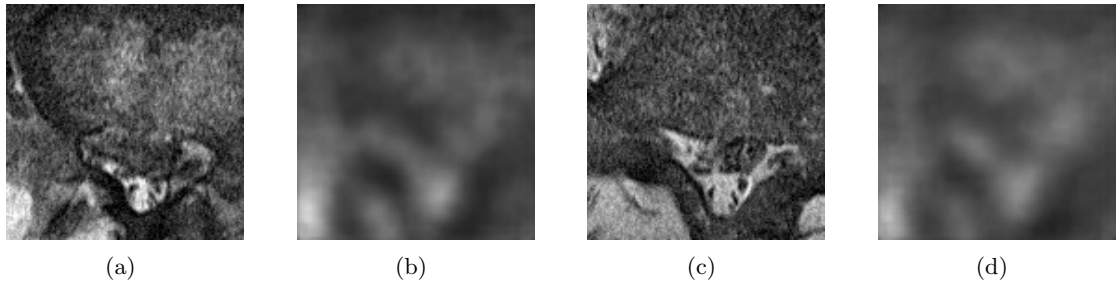

Figure 5: Image reconstruction examples by the autoencoder. Figures 5(a) and 5(c): 2 out of 4 slices of the original 3D image. Figures 5(b) and 5(d): Reconstructed image slices.

## 4. Discussion

While the influence of MRI scans on surgical decisions for LSS was previously unclear, our results quantitatively confirm the importance of medical imaging in LSS diagnosis and treatment planning. We started by effectively modeling surgical decision-making for lumbar spine stenosis through binary classifiers, on the sole basis of manually-assessed radiological features. To reduce human bias and errors in the selection and calculation of features, we developed an automatic pipeline (fig. 4) to work on raw MRI scans. To the best of our knowledge these are the first and initial steps towards benchmarking LSS. Supervised (CNN) and semi-supervised (convolutional autoencoders) deep learning algorithms were trained on the transformed images and accuracies around 70% on surgical planning were achieved. Compared to the results with the numerical dataset, the differences in accuracy (of about 15%) can be justified by the modest number of MRI scans. We are confident that further systematic efforts aimed at enlarging the image catalog could significantly improve the classification results and thus patient outcome.

## Appendix A. Supplementary Material

### A.1 Dataset Features

The numerical dataset employed for the experiments includes 6 quantitative and 16 qualitative variables. In the following a detailed description of the features used in the classifier experiments in section 2 is given.

1. `SegDisc`: discrete integer variable taking values in $\{1, 2, 3, 4, 5\}$. It describes the structure of the discus, where 1 indicates a homogeneous, white on MRI T2 images, and 5 indicates a black, collapsed, inhomogeneous discus.

2. `SegModic`: discrete integer variable taking values in $\{1, 2, 3\}$, it identifies changes in the lumbar vertebral marrow. 1 represents bone marrow edema and inflammation, 2 represents normal red hematopoietic bone marrow conversion into yellow fatty marrow as a result of marrow ischemia, 3 represents subchondral bony sclerosis.

3. `SegListhesis`: binary variable, indicates the presence of listhesis (displacement of the vertebra).

4. `SegOsteoArthRight`: discrete integer variable taking values in $\{0, 1, 2, 3\}$. It describes the degeneration of the right facet joint: 0 indicates a normal joint, 1 indicated narrowing, 2 indicates narrowing with sclerosis and hypertrophy, 3 indicates narrowing, sclerosis and the presence of osteophytes.

5. `SegOsteoArthLeft`: same as the previous one, but this time the left facet joint is described.

6. `SegFlavumRight`: binary variable, indicates hypertrophy of the right ligamentum flavum.

7. `SegFlavumLeft`: binary variable, indicates hypertrophy of the left ligamentum flavum.

8. `SegFlavumThickRight`: quantitative integer variable measuring the thickness in mm of the right ligamentum flavum.

9. `SegFlavumThickLeft`: quantitative integer variable measuring the thickness in mm of the left ligamentum flavum.

10. `SegLipomathosis`: discrete integer variable taking values in $\{0, 1, 2, 3\}$, where 0 indicates a normal amount of epidural fat and 3 indicates severe epidural fat overgrowth.

11. `SegCentralZone`: discrete integer variable taking values in $\{0, 1, 2, 3\}$, it describes the compromise of the central zone of the vertebra. 0 corresponds to no compromise of the central zone while 3 to a severe compromise (affecting the 2/3 of its normal size).

12. `SegFluidSign`: discrete integer variable taking values in $\{0, 1, 2, 3, 4, 5, 6, 7\}$, it indicates stenosis by assessing the relation from the fluid to the cauda equina. The grading goes from a minor stenosis (value 0) to a extreme stenosis (value 7), with all the intermediate values in the middle.

13. `SegRecessRight`: discrete integer variable taking values in $\{0, 1, 2, 3\}$, it assesses the nerve root compression in the right lateral recess. Grade 0 identifies absence of compression while grade 3 corresponds to severe nerve root compression within the lateral recess with obliteration of cerebrospinal fluid from the recess.

14. `SegRecessLeft`: same as the previous one, but this time the left lateral recess is described.

15. `SegRootRight`: discrete integer variable taking values in $\{0, 1, 2, 3\}$, it describes the right foraminal nerve root impingement. Grade 0 corresponds to the absence of impingement, grade 3 indicates compression between disk material and the wall of the spinal canal, such that it may appear flattened or indistinguishable from disk material.

16. `SegRootLeft`: same as the previous one, but the impingement considered here is the one on the left nerve root impingement.

17. `SegForamenRight`: discrete integer variable taking values in $\{0, 1, 2, 3\}$, grading the compromise of the right foraminal zone from absence of compromise (grade 0) to a compromise affecting more than the 2/3 of its normal size.

18. `SegForamenLeft`: same as the previous one, but the left foraminal zone is considered here.

19. `SegAPDIam`: quantitative variable, equal to length in mm of the anterior-posterior diameter of the dural sac.

20. `SegCSArea`: quantitative variable, equal to the cross-sectional area in mm$^2$ of the dural sac.

21. `SegLRRight`: quantitative variable, equal to the depth in mm of the right lateral recess.

22. `SegLRLeft`: quantitative variable, equal to the depth in mm of the left lateral recess.

### A.2 Learning Curves

We report here the learning curves that describe the training of the deep learning algorithms. In figure 6 the evolution of accuracy (6(a)) and cost function (6(b)) during 100 epochs of training of the convolutional neural networks are shown (a moving average is plotted for clarity). In figure 7 we exhibit the trend of the mean cost function during the training of the autoencoder. Recall the autoencoder is trained to minimize the tensor:

$$\mathbf{J} = (\mathbf{X}_{\text{orig}} - \mathbf{X}_{\text{reconstr}})^2, \tag{1}$$

an example of successful image reconstruction is given in figure 8. The plot in figure 7 depicts the scalar function $\mathbb{E}[\mathbf{J}]$ during the 100 epochs of training.

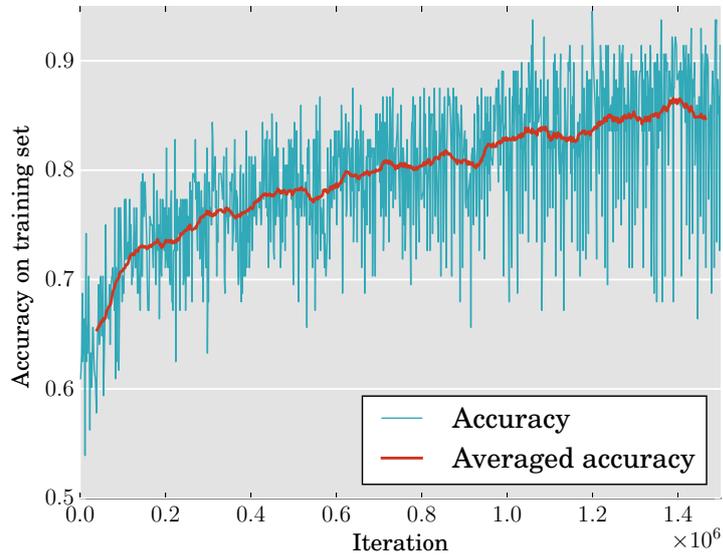

(a) CNN: accuracy

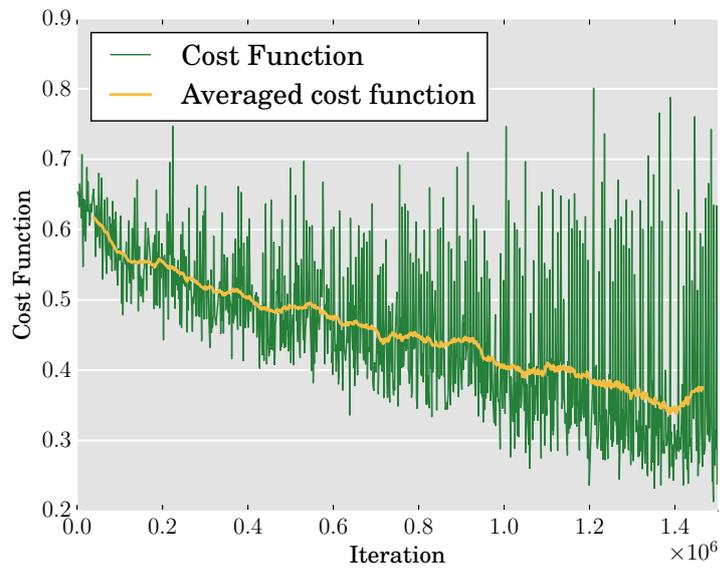

(b) CNN: cost function

Figure 6: Learning curves describing the training of the convolutional neural network (CNN). Figure 6(a) shows the accuracy on the training set, while figure 6(b) shows the cost function during training. Moving averages are plotted for clarity.

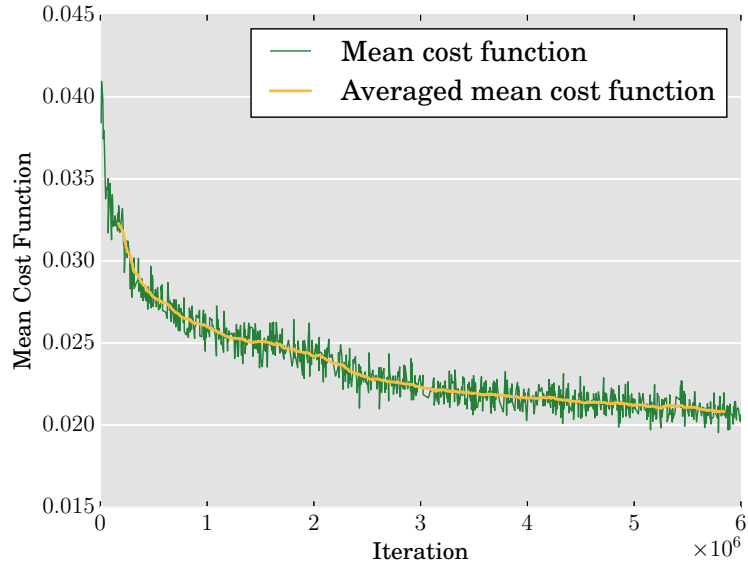

Figure 7: Mean cost function ($\mathbb{E}[\mathbf{J}(\boldsymbol{\theta})]$, see eq. 1) during the training of the convolutional autoencoder. A moving average is plotted for clarity.

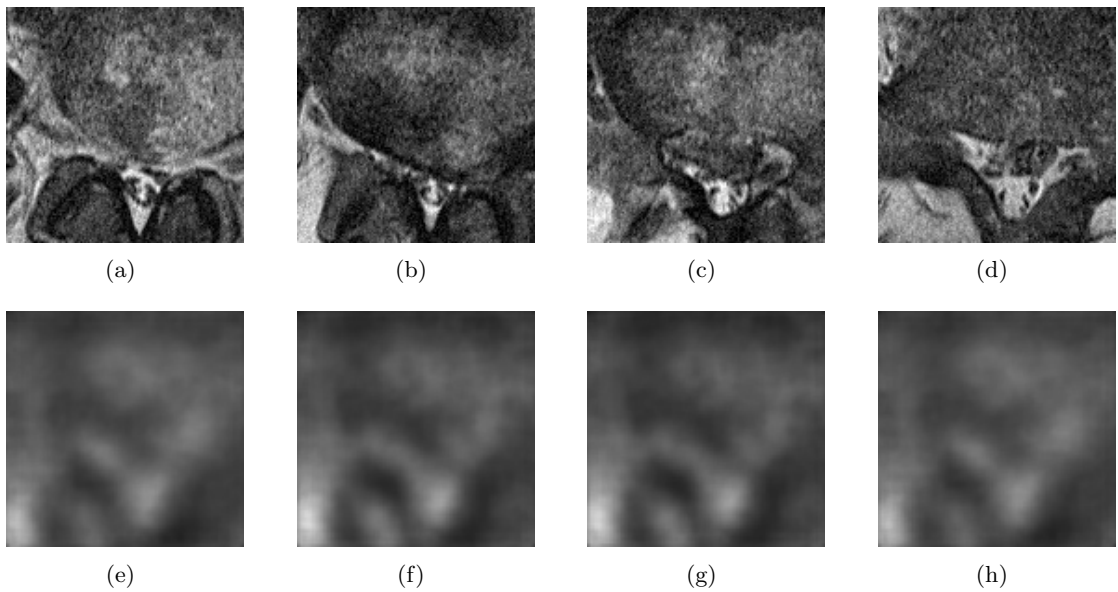

Figure 8: Image reconstruction examples by the autoencoder. Figures 8(a) to 8(d): Original 3D image slices 1 to 4. Figures 8(e) to 8(h): Reconstructed images from slices 1 to 4.

4